\documentclass[10pt]{article}
\usepackage{geometry}
\usepackage{cite}
\usepackage{amssymb,amsmath}
\usepackage{wrapfig}
\usepackage{graphicx}
\usepackage[small]{caption}
\usepackage[title]{appendix}
\usepackage{authblk}
\usepackage{abstract}

\begin{document} 
\title{A Nonlocality Anomaly and Extended Semiquantum Games}
\author{Yiruo Lin \thanks{yiruolin@tamu.edu, linyiruo1@gmail.com}}
\affil{\textit{Department of Computer Science \& Engineering, Texas A\&M University}}
\date{}
\maketitle

\begin{abstract}
\normalsize
A nonlocality anomaly in which a partially entangled state can outperform a maximally entangled state in a task exploiting nonlocality and several ways to remove the anomaly are discussed. A necessary condition for the anomaly to occur is given in terms of joint probabilities of local measurements. By extending semiquantum games to include classical communication, the anomaly is shown to be removed with respect to semiquantum probabilities. 
\end{abstract}

\section{Introduction} \label{intro}

Quantum mechanics has predicted surprisingly many puzzling phenomena alien to the notions of a classical world. Quantum nonlocality, arising from quantum entanglement\cite{caveat} and perhaps the most famous and profound puzzle, suggests a nonseparable world in which action on one part of the world can affect other parts that have space-like separations from it. People are vigorously exploiting the nonclassical nature of quantum mechanics for applications in quantum computation and quantum information processing with the intriguing prospect of superior performance to classic systems \cite{Chuang}. In particular, entanglement and nonlocality have witnessed intensive interest in the physics community and are being actively exploited as useful resources for applications in information processing such as cryptography, dense coding, teleportation, and communication complexity \cite{RMP-nonlocal,QE,steering}. \\

A quantum state is entangled if it can not be written as a mixture of product states. Correlations of measurement outcomes of spatially separated parties are said to be nonlocal if they can not be reproduced by shared randomness that gives rise to classical correlations. For a pure quantum state, entanglement guarantees nonlocality in the sense of violating at least one Bell inequality \cite{Bell_ineq}. However, situation becomes complicated for mixed states (whose density matrix can be written as a convex sum of pure quantum states) where quantum nonlocality due to entanglement is mixed with classical correlations and sometimes becomes hidden from measurements\cite{Werner,POVM_Werner}. Even more surprising is the counter intuitive observation that there exist anomalous cases where maximal nonlocality is not realizable by maximally entangled states, but by partially entangled states. This so called anomaly of nonlocality \cite{anomaly_rev} signals a gap in our understanding of the relation between entanglement and nonlocality. \\

In this article, we discuss a necessary condition of the anomaly and show how this condition is removed in extended semiquantum games. The paper is organized as follows. In section \ref{anomaly}, we give a brief review of the anomaly and some ways to remove it. We then point out a particular aspect of the anomaly. In section \ref{semiquan}, we review semiquantum games. In section \ref{extended}, we extend the semiquantum games to include classical communication in the set of free operations and show how they remove the anomaly. Finally, we conclude with discussions on the future work in section \ref{disc}.

\section{the Anomaly of Nonlocality} \label{anomaly}
The anomaly can take various manifestations depending on quantum resources, schemes, tasks and respective measures of nonlocality. It is regarded to occur whenever a less entangled state can give rise to more nonlocality than a more entangled state. In all reported cases (to the best of the author's knowledge), entanglement is measured by entanglement entropy and the quantum states in comparison are pure states. The entanglement entropy for a pure bipartite state $|\Psi\rangle_{AB}$ is $S(\rho_A)=S(\rho_B)=-\mathrm{Tr}(\rho_A\mathrm{log}\rho_A)=-\mathrm{Tr}(\rho_B\mathrm{log}\rho_B)$ with $\rho_A=\mathrm{Tr}_B(|\Psi\rangle\langle\Psi|_{AB})$ and $\rho_B=\mathrm{Tr}_A(|\Psi\rangle\langle\Psi|_{AB})$ where $\mathrm{Tr}_B(\cdot)$ is partial trace over subsystem $B$ and similarly for $\mathrm{Tr}_A(\cdot)$. With respect to the entanglement entropy, the maximally entangled states of local quantum dimension $n$ (such as singlet and triplet states in two-qubit systems) take the form $|\Psi\rangle_{\mathrm{max}}=1/\sqrt{n}\sum_{i=1}^n|i\rangle_A|i\rangle_B$ in Schmidt decomposition where all Schmidt coefficients are of equal magnitude $1/\sqrt{n}$ and partially entangled states correspond to some Schmidt coefficients taking unequal weights. On the other hand, various measures of nonlocality are adopted depending on the schemes for detecting nonlocality along with the specific tasks. The most popular measure is based on the violation of a Bell inequality, which is an inequality in terms of outcome correlations of local measurements on spatially separated systems and can be written for bipartite correlations as $\sum_{a,b,x,y}\xi_{abxy}P(a,b|x,y)\leq0$ (where $P(a,b|x,y)$ is the joint probability of obtaining result $a$ and $b$ for Alice and Bob conditional on their measurement inputs $x$ and $y$ respectively, $\xi_{abxy}$ is coefficient). A Bell inequality is satisfied by all classical correlations (local correlations admitting shared randomness parameterized by local hidden variables $\lambda$: $P_c(a,b|x,y)=\sum_\lambda P(\lambda)P(a|x,\lambda)P(b|y,\lambda)$ for bipartite classical correlations $P_c(a,b|x,y)$) thanks to the linearity of the inequality and the convex polytope formed by the set of classical correlations. Geometrically, a Bell inequality represents a hyperplane in probability space delineating nonlocal correlations from local correlations. The amount of violation of a Bell inequality may be viewed as distance from such a surface and hence an intuitive measure to quantify the degree of nonlocality. \\

For the simplest Bell inequality, the CHSH inequality of bipartite correlations of two measurement settings and two measurement outcomes per party (i.e., each party performs two different measurements and each measurement has two outcomes), the anomaly doesn't occur for pure bipartite quantum states of qubits. However, it has been observed for a variety of other Bell inequalities. For example (cf. \cite{anomaly_rev}), the detection loophole efficiency for the CHSH inequality corresponding to a modified CHSH inequality is better for some partially entangled qubit states than for maximally entangled counterpart. Beyond the simplest measurement settings, the anomaly is also observed for the next simplest case of a two-qutrit system probed with two measurement settings and three measurement outcomes per party. The corresponding Bell inequality - the CGLMP inequality, is maximally violated by partially entangled qutrit states instead of maximally entangled states. It is interesting to note that for the same quantum system and measurement settings, the anomaly can also be observed with respect to a different nonlocality measure based on Kullback-Leibler(K-L) distance that quantifies the statistical distance between the correlations in measurement outcomes and classical correlations. In this case, the optimal nonlocality is again achieved by partially entangled states, but different from those maximally violating the CGLMP inequality. Furthermore, for another measure based on rate of secret key extraction that exploits nonlocality, the best performance is achieved by yet another set of partially entangled states. That the maximal value of various nonlocality measures is taken by different partially entangled states shows a disagreement among these measures in quantifying nonlocality. Related work reported similar anomaly in violating the CGLMP inequality including measurement outcomes larger than 3 \cite{CGLMP_d} which is further generalized for more measurement outcomes together with a detailed comparison of K-L measure (based on the above mentioned K-L distance) with Bell measure (i.e., violation of a Bell inequality, here the CGLMP inequality) \cite{CGLMP_general}. Anomaly associated with other Bell inequalities such as a set of modified CHSH inequalities \cite{modCHSH}, and I3322 Bell inequality \cite{I3322} was later found and further explored for interesting applications. \\

The anomaly has also been observed in various other scenarios besides Bell inequalities. For instance (\cite{anomaly_rev} and citations therein), in simulating entangled states with nonlocal resources, more resource is required for partially entangled states than maximally entangled states; Hardy's proof of nonlocality works for almost any entangled state of qubits but not for maximally entangled states. More intricacy involving entanglement and nonlocality was explored in \cite{random} where randomness of local measurement outcomes (which is useful for cryptography applications) on entangled states was shown to behave rather independently of entanglement and Bell nonlocality, a surprising counterexample to the usual wisdom of more nonlocality, better performance of entangled states. Furthermore, if the randomness is to be regarded as a measure of nonlocality, the finding shows another case of disagreement between different nonlocality measures, i.e., between randomness measure and Bell nonlocality measure. \\

The work reviewed hitherto on the anomaly reveals rich physics underlying the relation between entanglement and nonlocality. In particular, the anomaly depends on the specific aspect of nonlocality and the corresponding nonlocality measure in the sense that different entangled states, even nearly unentangled state \cite{random} can show maximal nonlocality. \\

The physical reason underlying the anomaly has been  discussed in several pieces of work. The distinction in operational meaning between entanglement entropy and nonlocality measures was pointed out in \cite{anomaly_rev}. It was argued that since the measures for the two entities quantify performance of different quantum resources with respect to different free operations, there is no a priori reason to identify entanglement with nonlocality and hence the anomaly is not necessarily anomalous despite its intuitive appearance. It has later been further studied in more details in the framework of resource theory (for a related review, see \cite{QRT}) in which the anomaly was shown to be of no paradoxical nature \cite{Schmid-interplay}. Roughly speaking, some pairs of convertible entangled states become inconvertible under free operations argued to be more appropriate in regard to characterizing nonlocality, as the set of free operations is more restricted compared to that for entanglement. Specifically, classical communication (\textbf{CC}) can not create entangled states from separable states and is  thus a legitimate free operation whereas CC is not allowed for manipulating nonlocality since it can create nonlocal correlations. Maximally and partially entangled states are inconvertible under local operations and shared randomness (\textbf{LOSR}), the legitimate free operations under which to characterize nonlocality. Hence, there should be no a priori constraint on the relative amount of nonlocality associated with the maximally and the partially entangled states. (The amount of nonlocality is expected to be nonincreasing only under conversion by local free operations. There is no a priori requirement on the relative nonlocality exhibited by a pair of entangled states that are inconvertible to each other. ) \\

In the author's opinion, examining convertibility between entangled states under LOSR as done in \cite{Schmid-interplay} is actually of no direct relevance a priori to resolving the anomaly. This is due to the simple fact that LOSR on entangled states and LOSR on quantum boxes \cite{box} (or equivalently joint probabilities of local measurement outcomes on a shared quantum system) are independent sets of operations acting on distinct objects. To assess the anomaly in the framework of resource theory, it is more intuitive to compare  the convertibility of quantum boxes with that of the corresponding entangled states. In order for the resource theory to be relevant, any pair of entangled states in consideration should be ordered, i.e., one of them can be converted to the other, since otherwise there would be no a priori physical relevance with respect to the relative magnitudes of the entanglement measures. It is clear here that the proper operation for entanglement should be LOCC (or any free operations enclosing it, e.g., separable operations) under which maximally entangled states can be converted to partially entangled states, but not the other way around. Whenever an anomaly occurs, the corresponding pair of boxes must be either incomparable or be ordered differently (i.e., the box by a partially entangled state is convertible to that by a maximally entangled state). \\

Another point of view \cite{Bell_entanglement} is to simply dismiss the anomaly by including most general free operations (including CC) in the allowed set of free operations and unifying Bell nonlocality with entanglement in the framework of dynamical resource theory. This argument, however, doesn't seem to answer directly the question whether a maximally entangled state can always produce more nonlocality than any of the partially entangled states. It merely guarantees (in a somewhat trivial sense) that the former can produce at least the same nonlocality as any of the latter. \\

A different line of work \cite{random_meas1, random_meas2, random_meas3} examined a particular type of nonlocality measure based on the probability of violating a Bell inequality by some random local measurements, which can remove the anomaly in some scenarios for which the highest probability is achieved by a maximally entangled state.  \\

We note here a simple aspect shared by all the anomalous cases: the nonlocal quantum boxes generated by a maximally entangled state do not enclose the set of boxes generated by a partially entangled state when an anomaly occurs. In this aspect, a way to remove the anomaly is to find measurement correlations with respect to which a maximally entangled state generates a set that strictly encloses the set by any of the partially entangled states. We can then define nonlocality measures on the measurement correlations for which a maximally entangled state outperforms any partially entangled state, but not the other way around. For this purpose, we need to look for measurement correlations generated by resources beyond boxes. \\

\section{Semiquantum Games} \label{semiquan}

The anomaly may be viewed as due to degradation of nonclassicality in an entangled state during its transformation to a box by local measurements \cite{Schmid-LOSR}. So a natural resolution is to preserve the nonclassicality during its transformations to other types of resources. In \cite{Schmid-LOSR}, the effect on the nonclassicality was examined in a resource during its transformation into a different type and it was found that the nonclassicality of any resource can be fully preserved when transformed to one of the higher type \cite{type}. An example is the well known semiquantum games \cite{semiquantum} in which all mixed entangled states, including those which admit local hidden variable models, can display nonlocality. \\

It is instructive to briefly go through the relevant argument in \cite{Schmid-LOSR}. All discussions in \cite{Schmid-LOSR} were restricted to LOSR under which the type of semiquantum resources (i.e., resources with quantum input and classical output) is the highest, namely for any resource of any type there always exists a member of semiquantum resources interconvertible to it (see Figure 4 of \cite{Schmid-LOSR} for an illustration). The order between any two resources of a given type is preserved in their optimal performances in all games with resources of a higher type. Namely, if one resource can be converted to the other under LOSR, then its optimal performance $\omega_{G_T}$ in any game $G_T$ with resources of a higher type $T$ is no worse than the other, and vice versa: $R\xrightarrow[\text{}]{\text{LOSR}}R'\Longleftrightarrow\omega_{G_T}(R)\geq\omega_{G_T}(R')$ for all $G_T$. The optimal performance of a resource $R$ in a game with a given resource type $T$ is defined by the best performance among resources of type $T$ that are convertible under LOSR from $R$. The performance is quantified by a real number that is mapped linearly according to the game from a set of joint probabilities associated to a resource: the larger the number, the better the performance. In a nutshell, the preservation of order between two resources in every game of a higher resource type is mainly due to convexity of LOSR and linearity of games. It is trivial to show $R\xrightarrow[\text{}]{\text{LOSR}}R'\Longrightarrow\omega_{G_T}(R)\geq\omega_{G_T}(R')$. To prove the converse, note that if $R$ can not be converted to $R'$, or any resource of type T convertible from $R'$, then by the convexity of LOSR and the linearity of games, there always exists a game with the corresponding hyperplane separating the set of probabilities by $R'$ from the convex hull of probabilities by all resources convertible from $R$.  Such a game results in better performance by $R'$ and we reach a contradiction. Hence, any resource $\mathcal{E}_T$ convertible from $R'$ can also be converted from $R$. Since there exists $\mathcal{E}_T$ such that $\mathcal{E}_T\xrightarrow[\text{}]{\text{LOSR}}R'$ ($\because$ $\mathcal{E}_T$ is of higher type than $R'$), we prove by transitivity that $\omega_{G_T}(R)\geq\omega_{G_T}(R')\Longrightarrow R\xrightarrow[\text{}]{\text{LOSR}}R'$.   \\

For a mixed entangled state admitting a hidden local variable model, any box generated by it can be obtained by a separable state. In other words, the set of boxes generated by it is enclosed by the set generated by the union of all separable states. Hence no Bell inequality can be violated. When the mixed entangled state is put into a semiquantum game, since no separable state can be converted to an entangled state under LOSR, by the above argument, there exists a semiquantum game in which the performance of the mixed entangled state is superior to any of the separable states. In this sense, the hidden nonlocality is displayed by the semiquantum game. \\

\section{Extended Semiquantum Games and the Anomaly} \label{extended}

We can actually generalize readily the above proof on the preservation of resource order to entangled states ordered under LOCC. A maximally entangled state is higher in order than a partially entangled state under LOCC. If we extend the free operations from LOSR to LOCC in the definition of optimal performance of a resource, the order of a maximally entangled state over a partially entangled state is preserved in a semiquantum game. The operations under which resource type is ordered can be either LOSR as adopted in \cite{Schmid-LOSR} or LOCC. Since LOCC encloses LOSR, semiquantum resource is still the highest type under LOCC. Needless to say, replacing LOSR by LOCC changes order of some resources, for instance, the order of boxes becomes trivial since classical communication after local measurements can access any nonlocal box. Nevertheless, the preservation of order in entangled states in extended semiquantum games involving LOCC remains valid as all the argument in the previous paragraph goes through with LOCC. \\

Since no partially entangled states can be converted to a maximally entangled state under LOCC, there must exist a semiquantum game in which the optimal performance associated to any partially entangled state with respect to LOCC is worse than the maximally entangled state. Equivalently, this means that there exists a set of joint probabilities (with quantum inputs and classical outputs) associated to  a maximally entangled state that is outside the convex hull of joint probabilities associated to any partially entangled state. Thus, the anomaly is removed in the sense that there exists a task with optimal performance by a maximally entangled state unachievable by any partially entangled state, but not the other way around.\\

\section{Discussions} \label{disc}

A characteristics of the anomaly of nonlocality is that the set of boxes generated by a maximally entangled state doesn't enclose the set of boxes by a partially entangled state. By extending the semiquantum games to include classical communication,  we show that the set of semiquantum probabilities associated to a maximally entangled state strictly enclose the probabilities of any partially entangled state. Hence the extended semiquantum games remove the anomaly: the performance of a maximally entangled state strictly encloses that of any partially entangled state.  \\

Our discussions on the anomaly are closely related to the probability distribution over probability space. This is reminiscent of the line of work \cite{random_meas1, random_meas2, random_meas3} on removing the anomaly mentioned in part \ref{anomaly}, which explored the probability distribution generated by local random measurements. The statistical weight associated to joint measurement probabilities lying on each side of the hyperplane in the probability space corresponding to a Bell inequality gives rise to the probability of successfully violating the Bell inequality. In light of those work, we may alternatively consider the semiquantum games in its original form (i.e., involving only LOSR) and calculate, for instance, the probability of semiquantum probabilities generated by random semiquantum measurements on a partially entangled state to be reached by a maximally entangled state (and vice versa). \\

It is worth noting a related work which explored the semiquantum games \cite{semi}. The anomaly is removed by a unification of semiquantum nonlocality with entanglement in terms of robustness measures. We believe there are more aspects to explore in the semiquantum games for a better understanding of the nonlocality anomaly. \\

\section*{Acknowledgement}

The author thanks the support from the Department of Computer Science \& Engineering, Texas A\&M University.


\begin{thebibliography}{99}
\bibitem{box} A box is a process with only classical inputs and outputs realizable by a common source shared among parties who can perform local measurements on it such as in a Bell inequality test.
\bibitem{type} Both input and output type of a local party sharing the quantum resource can be trivial, classical or quantum. Trivial refers to no input/output, classical corresponds to classical variables parameterizing e.g. measurement settings in a Bell test, quantum involves quantum states. For example, an entangled state is a quantum resource with trivial input and quantum output. For more details, see \cite{Schmid-LOSR}.
\end{thebibliography}
\end{document}